# Next-Gen Education: Enhancing AI for Microlearning

**Introduction**

According to a report by Times Higher Education, a British magazine, class attendance at US universities has reached a historic low in the post-Covid era [4]. Academics argue that the lockdown measures imposed due to COVID-19 have led to the normalization of students skipping classes or opting for remote attendance [7]. Our observations indicate that class attendance decreases to 40% within the first month of the semester, consistent with statements and releases from The White House briefing room regarding state-specific actions to increase student attendance [13]. Another popular strategy is to implement graded classroom activities, resulting in a 10%-15% increase in attendance. However, an increasing trend is that many students opt not to attend classes in person, instead favoring recorded lectures and lecture notes as part of their assessment preparation.

Undoubtedly, video lectures and lecture notes are more effective for learning, as they allow students to proceed at their own pace and review content repeatedly to enhance understanding. However, we face challenges that surpass our initial expectations. Students often show disinterest in watching recorded lectures lasting 50 to 75 minutes, with longer lecture videos leading to higher dropout rates [17]. Traditional educational materials often fail to meet the specific needs of online learners. Research also indicates that students need non-sequential, step-by-step access to content without manually searching through the content to find specific information [17], [6], [30]. As a result, students often rely only on lecture slides for studying, leading to many questions during office hours. This creates challenges for large classes and impacts students' performance. To address these concerns and overcome barriers to the learning process, we propose integrating microlearning activities with traditional materials. By offering bite-sized, easily digestible content, this flexible approach empowers educators to cater to the specific needs of online learners, fostering a greater sense of control and efficacy in the online learning environment.

Microlearning has garnered significant attention in education due to its effectiveness in engaging students and enhancing their learning outcomes [23], [33], [22]. This approach involves breaking down complex topics into small, manageable chunks, making it easier for students to absorb and retain. It aligns well with the needs of modern learners, who often face distractions and have shorter attention spans [19]. Studies have shown that microlearning can increase retention rates and provide more personalized learning experiences [20]. Microlearning supports various delivery forms, including videos, interactive quizzes, and flashcards, making learning more accessible and flexible for students. Moreover, microlearning can effectively bridge the gap between theoretical knowledge and practical application by providing content directly applicable to real-world scenarios [1], [18]. This method focuses on delivering specific, actionable information that learners can immediately use in their daily tasks or professional activities.

Core computer science courses require a solid grasp of algorithms, programming logic, complex computations, and the principles of computer systems. These subjects necessitate focused attention from students. Educators play a crucial role in incorporating real-world scenarios into course materials to elucidate logical concepts or computational theories, making students feel empowered and integral to the learning process. For instance, when teaching propositional logic, the implication "$p \rightarrow q$" ($p$ implies $q$) might be illustrated using everyday examples like "If it is raining, then we take an umbrella." Adopting microlearning as a structured approach could significantly enhance early computer science education, enabling students to solidify fundamental principles from the start of their studies.

Creating microlearning materials can be time-consuming for educators [42]. However, Generative AI like ChatGPT [25] can streamline this process by generating personalized summaries, flashcards, and quizzes tailored to specific subjects. Generative AI has become a prevalent tool for content creation across various sectors today. Daniel examined the impact of generative AI on creating learning videos with synthetic virtual instructors, finding no significant differences between AI-generated and traditionally produced videos in learning gains and experience, suggesting AI-generated videos could effectively substitute traditional methods in online education [21].

This study explores the potential of AI-driven tools, specifically ChatGPT, to automate the creation of microlearning materials and evaluate their impact on educational outcomes. To provide a systematic foundation, this paper addresses two key research questions that guide the investigation: 1) *How do students perceive the effectiveness of AI-generated microlearning materials in supporting their learning and engagement?* 2) *And how accurate is the content produced by large language models (LLMs) for microlearning?*

The rest of this paper is structured as follows: The next section presents a literature review contextualizing the research within existing studies on AI in education. The methodology section outlines the processes for implementing AI-driven microlearning tools and evaluating their effectiveness and accuracy. The experiment and result section provides insights derived from quantitative and qualitative data from two junior-level computer science courses to address research questions. The discussion section interprets these findings, offering recommendations and implications for future applications, and the conclusion summarizes the key contributions.

**Literature Review**

Integrating artificial intelligence (AI) into educational technologies has revolutionized how content is delivered and personalized, making it more adaptive to individual learners' needs. In this context, microlearning has emerged as a key strategy for delivering concise, targeted educational materials, with AI playing a pivotal role in its evolution. This section reviews recent works highlighting the potential and challenges of AI-driven microlearning.

Larrondo [34] demonstrated the effectiveness of fine-tuned large language models (LLMs) in providing automated feedback for complex engineering problem-solving tasks. Their work showcased how LLMs could deliver personalized guidance, significantly improving the learning process. Similarly, Ivanov and Soliman [35] highlighted the potential of ChatGPT to generate modular educational content tailored to individual learners. While their research focused on applications in non-STEM fields, it provides a foundation for adapting similar technologies to

STEM education, aligning with the objectives of microlearning to enhance accessibility and engagement.

Sridhar [36] extended this approach by leveraging GPT-4 to generate modular learning objectives aligned with Bloom's taxonomy. Their findings underscore the importance of structured, goal-oriented content in educational technologies. This parallels the objectives of microlearning, which seeks to deliver concise yet effective learning experiences catering to cognitive learning goals.

Microlearning's emphasis on delivering brief, focused content is supported by research demonstrating its impact on learner engagement and knowledge retention. Manning [37] investigated the effectiveness of Bite-Sized Teaching (BST) in postgraduate medical education, showing that short, targeted content significantly improved knowledge acquisition and learner satisfaction. Cheung [38] evaluated ChatGPT's capability to generate modular multiple-choice questions for medical education. The study revealed that AI-generated content could achieve high scalability and efficiency, emphasizing the role of modular learning resources in modern education.

The cognitive principles underlying microlearning are grounded in cognitive load theory, which emphasizes segmenting content into manageable units to optimize learning. Hug [39] and Mayer [40] discussed how breaking down complex information into smaller chunks reduces cognitive load, making learning more effective. These foundational principles align with microlearning's structuring of educational content for better learner outcomes. Bartram [41] further validated the effectiveness of such strategies by demonstrating how bite-sized simulations in medical training enhanced engagement and reduced cognitive overload.

Despite its potential, the integration of AI in microlearning faces challenges. Issues such as bias in AI-generated content, ethical concerns, and over-reliance on automation remain critical areas of discussion. Ivanov and Soliman [35] cautioned against the lack of depth and critical analysis in AI-generated materials, which could lead to surface-level understanding if not carefully moderated. Addressing these challenges requires rigorous human oversight and iterative refinement of AI tools to ensure the accuracy and relevance of the content.

**Method**

This section introduces a set of microlearning elements created using the automated method outlined in this paper. Furthermore, we outline the steps in the proposed methodology for developing the intended microlearning components. The latter part of this section outlines the methodology used to evaluate the effectiveness and accuracy of AI-generated microlearning.

*Microlearning Elements*

We target microlearning elements frequently used in education. The following subsections provide a detailed overview of the targeted microlearning elements in this study.

- *Interactive Quizzes*, as digital assessments, play a crucial role in engaging learners. They actively involve students in the learning process, providing instant feedback, hints, and explanations. By incorporating various question types and multimedia elements, these quizzes can significantly enhance the learning experience, making it more dynamic and effective [30].

- *Digital Flashcards* is an electronic version of a traditional flashcard used for learning and memorization. Digital flashcards typically feature a question or prompt on one side and the answer on the other. They can be accessed and used on digital devices such as computers, tablets, and smartphones, often through dedicated apps or online platforms. Digital flashcards can include multimedia elements like images, audio, and videos to enhance the learning experience [30], [45].
- *Mini Lessons* are a brief, focused instructional session to teach a specific concept or skill. These sessions usually last 5 to 15 minutes and are commonly used to introduce a topic, reinforce previously learned material, or provide quick, targeted practice [46].
- *Scenario-based Learning* (SBL) is an instructional strategy that uses realistic scenarios to actively engage learners in problem-solving and critical thinking. This approach enables learners to apply their knowledge and skills in situations that resemble real life, thereby improving their comprehension and retention of the material [30], [43], [44].

*Creating Microlearning Elements*

By leveraging functionalities offered by the OpenAI tools [24], we aim to establish a pipeline for converting video lectures and lecture slide content into microlearning formats. The block diagram in Figure 1 outlines the steps involved in this process, which can be divided into three significant steps: 1) transcribing a given video into text, 2) refining the video transcript, and 3) generating microlearning elements with the refined transcript and the provided lecture slides. This section explains the systematic steps required to generate microlearning elements.

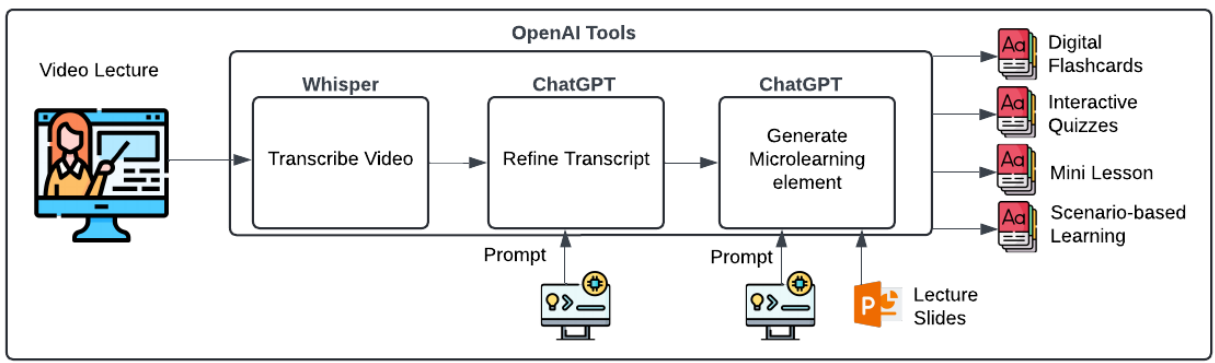

Figure 1: Process to create microlearning elements from lecture videos and slides

- *Transcribe Video*: Computer science courses comprise various materials such as lecture slides, notes, and practice problems. Our proposed system prioritizes using lecture videos as the primary input. This preference is due to the comprehensive nature of lecture videos, which offer structured explanations covering all topics thoroughly. Transcribing a lecture video can be done using any proficient video-to-text tool; here, we use Whisper

- [26], a machine learning model for speech recognition and transcription created by OpenAI.
- *Refine Transcript*: Lecture videos, often recorded in classrooms, may inadvertently capture incidental conversations and unstructured sentences. These, along with improper word use due to factors like incomplete audio recordings or instructor accents, necessitate the refinement of the transcript before generating microlearning content. We use ChatGPT [25] offered by OpenAI to automatically refine the transcript, ensuring the accuracy and clarity of the educational materials.
- *Generate Microlearning Elements*: This step involves developing microlearning content for each topic discussed in the lecture video. Utilizing the capabilities of AI tools like ChatGPT, we create these educational resources by inputting both the refined transcripts and the lecture slides. Including lecture slides along with the instructor's annotations is crucial because they contain pseudocode, formulas, and programming codes essential for crafting practical microlearning elements. Relying solely on the video lecture transcript would be inadequate for producing comprehensive microlearning materials such as programming quizzes or explanations of formulas.

It is imperative to conduct manual checks on the generated microlearning content to ensure its quality and accuracy before making it available to students. The generated content can be utilized in subsequent semesters or as training data to refine the AI tool's capabilities in generating different educational materials in future iterations.

*Prompt*

A prompt is an instruction given to a large language model (LLM), such as ChatGPT, enabling the tool to perform a specific task. The efficacy and relevance of an LLM's output are significantly influenced by how they are prompted [27]. Our approach uses prompts to instruct ChatGPT to refine a given transcript and create various microlearning elements.

*Experimental Setup*

This study evaluates the effectiveness and accuracy of AI-generated microlearning through a comprehensive approach. The study was carried out in two junior-level computer science courses at Pennsylvania State University, USA. We selected Discrete Mathematics and Programming Language Principles due to their contrasting nature—one focusing on abstract, theoretical concepts and the other on practical, hands-on problem-solving. This selection allowed for a thorough evaluation of AI-generated microlearning materials across different types of content and learning objectives.

Both courses, with a combined enrollment of 650 students, reflect large class sizes, catered to a diverse student population primarily consisting of junior-level undergraduates majoring in computer science or related disciplines. The courses were delivered in a hybrid format, offering students access to both in-person lectures and recorded sessions. This diverse student body and flexible delivery format provided a comprehensive testing ground for evaluating the effectiveness and accuracy of microlearning materials.

Microlearning materials, including interactive quizzes, digital flashcards, mini-lessons, and scenario-based exercises, were integrated into the coursework for both classes. However, the frequency of microlearning assessments varied, reflecting the different assessment strategies in the two courses. In Programming Language Principles, 12 microlearning assessments were administered weekly throughout the semester, ensuring consistent interaction with the AI-generated content. In contrast, Discrete Mathematics included only 6 microlearning assessments spread across the semester, allowing for periodic evaluation of the microlearning materials' effectiveness. This difference in assessment frequency enabled the study to explore the impact of varying levels of integration on student engagement and learning outcomes.

The courses adopted different policies to examine the role of grading. In Programming Language Principles, completion of microlearning activities was graded, providing an incentive for active participation. In discrete mathematics, these activities were not graded, which offers a comparison between voluntary engagement and incentivized participation.

**Experiment and Results**

We have thoroughly examined existing tools, demonstrating our commitment to research and planning. This process has allowed us to identify tools applicable to our project's objectives and discern how we may effectively integrate those tools. Given our emphasis on automatically generating microlearning components from recorded video lectures, our primary focus is exploring the functionalities offered by the OpenAI tools. Our project has achieved significant efficiency and time-saving benefits by developing a Python script integrating various APIs from different tools. The script autonomously takes a lecture video and utilizes Whisper OpenAI's API to generate a video transcript. Subsequently, the script employs ChatGPT's APIs to refine these transcripts and create various microlearning elements, such as flashcards, mini summaries, interactive quizzes, and scenario-based exercises[1].

*Case Study*

In this section, we present an experiment demonstrating our proposed approach. We selected an essential topic from core computer science and followed all the steps outlined in the methods section to produce the proposed microlearning elements on the given topic.

*Topic of the Case Study*: Understanding the basics of pointers in C programming language is a significant challenge for many computer science students, especially those new to C programming. This struggle can result in a disinterest in learning more advanced pointer arithmetic in the future. As educators and researchers, it's crucial that we acknowledge and address these challenges.

*Inputs to the Case Study*: 1) Lecture Video: The primary input to our experiment is a 22-minute video lecture. The instructor in the video is not a native English speaker and has an accent. 2) Lecture Slides: The second input is a PowerPoint presentation consisting of five slides. These slides primarily contain programming examples to explain how to declare and define a pointer, de-reference a pointer, pass a pointer as a function argument, and return a pointer from a function.

---

[1] GitHub Repository: https://github.com/Fatemeh-Rahbari/AI-Microlearning-Generator

*Transcribe Lecture Video*: We use Whisper OpenAI to transcribe the given lecture. We chose this tool because of its AI-powered transcription capabilities, robustness to noise, and ability to handle diverse accents and dialects. However, after generating the transcript from the video lecture, we reviewed it and found several grammatical errors, improper word usage, and passages that needed clarification in the transcript. Furthermore, some video segments were inaccurately transcribed due to poor recording quality. These transcript inaccuracies could lead to misunderstandings or misinterpretations of the lecture content while generating microlearning elements. We give a sample transcribed version of the video lecture in Table 1. These challenges with the unstructured content led us to utilize an AI tool to improve the transcript.

> So when you declare the pointers, so before initializing, it basically has now, you know, it doesn't have anything, okay? So a pointer is a variable that contents and memory address that we have saved several times in the last few classes. And source of confusion is star P and P. So that's the two different things. So when we say the P, we know that P is a pointer variable.

Table 1: Original Transcript

*Refine Transcript*: To improve the transcript, we selected GPT-4o, OpenAI's pre-trained transformer model used in ChatGPT, due to its high accuracy and robust API support. This tool allows us to clean and structure the transcript efficiently, providing advanced grammar and punctuation correction, which is crucial for refining transcripts. It enhances readability and ensures the transcript is error-free, ensuring the content is ready for microlearning component creation. Table 2 presents the used prompt and the generated refined version of the content in Table 1.

> **Prompt:** *Refine the above lecture transcript to ensure it is clear, accurate, and professional. Remove any transcription errors, filler words (e.g., 'um,' 'uh,' 'you know'), repetitive phrases, and irrelevant noise. Ensure the terminology aligns with the subject matter and maintain the speaker's intended meaning. If any context is unclear, make logical edits to improve clarity while keeping the academic tone intact.*
>
> Initially, a declared pointer without initialization does not point to any specific memory address. Remember, a pointer stores the address of a memory location. A common source of confusion is differentiating *p from p. p refers to the pointer variable itself, while *p accesses the value at the memory address the pointer is pointing to.

Table 2: Refined Transcript

*Creating Microlearning Elements*: We have selected GPT4o, OpenAI's model powering ChatGPT, for its advanced natural language generation capabilities to create microlearning elements. Its robust NLP capabilities for text analysis, including sentiment analysis, keyword extraction, and entity recognition, make it a versatile tool. We use the refined transcript and the lecture slides to produce all the microlearning elements detailed in the methods section. Table 3 presents a multiple-choice question from the generated quizzes, while Table 4 illustrates a generated scenario-based learning activity, along with the associated prompts used for their creation.

> **Prompt:** *Using the provided refined transcript and slides, create a set of quizzes to evaluate students' understanding of the material. Ensure the questions cover key concepts, technical terms, and examples discussed in the lecture. Align the difficulty level with the course content and learning objectives and include clear and concise answer options or solutions for each question. For multiple-choice questions, ensure plausible distractors are included alongside the correct answer. Emphasize critical thinking and application of concepts where appropriate.*
>
> **Question:** If you have int *p1, *p2; and p1 = p2;, what does this mean?
>
> A) p1 and p2 now point to the same memory address.
> B) p2 is copied into p1.
> C) p1 is the parent pointer of p2.
> D) p1 and p2 are no longer usable.

Table 3: A Generated Quiz

> **Prompt:** *List the lecture objectives in the given refined transcript and slides, design scenario-based activities for each identified objective that encourage students to apply the concepts and skills covered in the material. Create realistic, context-driven scenarios relevant to the subject matter, requiring students to analyze, problem-solve, or make decisions based on the knowledge gained from the lecture. Ensure the scenarios are engaging, clearly structured, and include detailed instructions.*
>
> **Objective:** Practical Application of Pointers
>
> **Scenario:** You're involved in optimizing a legacy system that processes large datasets stored in arrays. The current implementation uses extensive loop-based array indexing, which has been identified as a performance bottleneck.
>
> **Task:** Optimize the data processing by implementing pointer arithmetic to traverse and manipulate the dataset.
>
> **Scenario-Based Learning Activity:**
>
> *Introduction:* Recap the relationship between arrays and pointers, highlighting how an array name can be used as a pointer.
>
> *Hands-On Activity:* Implement a function to find the mean and standard deviation of dataset values using pointers.
>
> *Assessment:* Discuss how pointer arithmetic enhances data processing efficiency, particularly

> in large datasets, compared to traditional array indexing.

Table 4: Generated Scenario-based Learning

*Readability Score*

The text in the study materials must align with their readability expectations to engage the target audience in AI-based microlearning environments. With its focus on delivering concise and engaging educational content, AI tools significantly enhance microlearning. These tools, by adjusting the complexity of the material to the learners' needs, whether they are engaging in casual learning or in-depth research, play a crucial role in not just delivering but optimizing content delivery. We implemented an automatic method to calculate readability scores using the TextStat library [3]. The library uses the Flesch Reading Ease formula [9] to measure text complexity by analyzing various aspects of the text, with a significant focus on sentence length and word difficulty.

| **Microlearning Type** | **Flesch Reading Ease Score** | **Grade Levels** |
|---|---|---|
| Flashcards | 73.58 | 7th to 8th-grade reading level |
| Quizzes | 48.71 | High school to college graduate reading level |
| Scenario-based learning | 49.25 | High school to college reading level |
| Mini lessons | 47.49 | 10th to college reading level |

Table 5: Readability Score

Table 5 shows the readability score for each generated microlearning category: flashcards, quizzes, scenario-based learning (SBL), and mini-lessons. A higher Flesch Reading Ease score indicates lower reading complexity. The flashcards, with a Flesch Reading Ease score of 73.58, are the easiest to read. They contain only 47 difficult words, making them highly suitable for middle school students. Quizzes have a lower Flesch score than flashcards, indicating higher reading complexity. However, mini lessons received the lowest score (47.49), suggesting they are slightly more complex. Despite this, they are still within the comprehension range of high school to early college students. The score reflects denser information and longer sentence structures, not necessarily conceptual difficulty. This nuance is important for interpreting how learners might interact with the content across different formats. This adaptability makes them suitable for many students, from high school to early college. These metrics ensure that each type of material is appropriately matched to the intended audience's reading capabilities, facilitating effective content delivery and optimal learning outcomes.

*Experiment Results*

The tool was deployed throughout the semester in the Discrete Math and Programming Language Principles courses to evaluate its scalability and effectiveness. As shown in Table 6, the tool successfully processed a significant amount of recorded lecture videos, approximately 225 minutes for each microlearning assessment. These assessments included diverse components such as flashcards, mini summaries, scenario-based activities, and quizzes. Remarkably, the tool demonstrated efficiency by processing 150 to 225 minutes of video content and over 100 pages of lecture slides for each assignment within just 45 minutes, reassuring the audience about its capabilities to handle substantial instructional material in a short period.

At the end of the semester, students in both courses were asked to complete a survey evaluating the effectiveness and accuracy of the AI-generated materials. The study graded in Programming Language Principles resulted in significantly higher participation (279 responses) than Discrete Mathematics (81 responses), where the survey was voluntary. This survey data provided critical insights into student perceptions and learning outcomes.

| Course name | Number of students | Number of assessments | Avg. video length per assessment | Participants in Survey |
|---|---|---|---|---|
| Discrete Math | 269 | 6 | 150 - 225 minutes | 81 |
| Programming Language Principles | 381 | 12 | 150 minutes | 279 |

Table 6: Details of classes and assessments

*Addressing RQ1*

This evaluation assesses effectiveness across three key dimensions: learning efficiency, engagement, and value as a starting point, to evaluate the impact of AI-driven microlearning solutions. Specifically, learning efficiency is measured through three metrics: time efficiency, improved retention, and interactive learning.

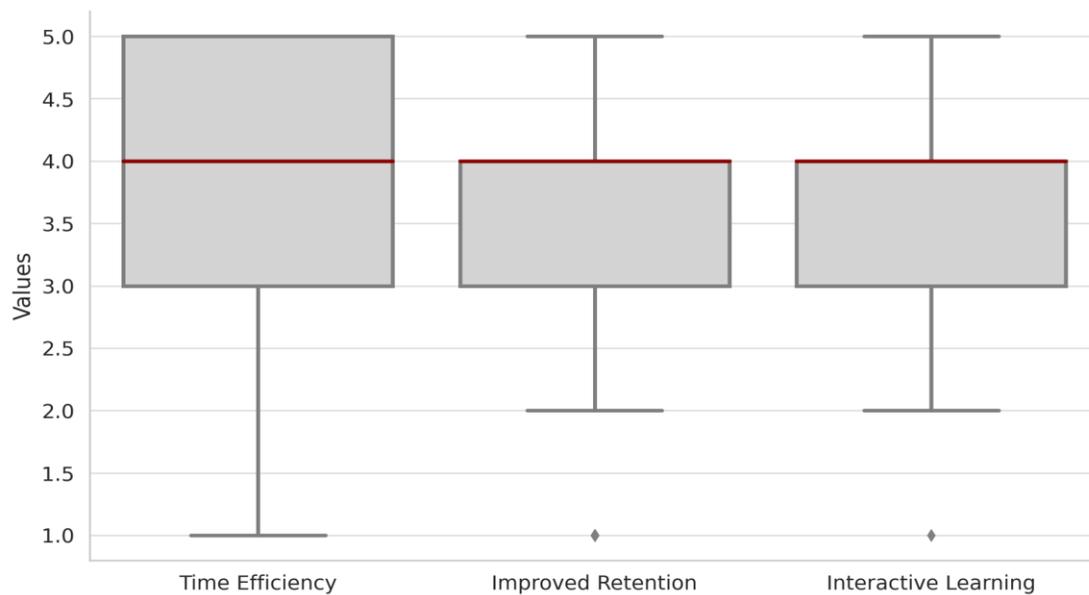

Figure 2: Students' feedback on the effectiveness of microlearning in enhancing learning efficiency

Based on survey outcomes, descriptive statistics in Figure 2 show that Time Efficiency has a mean of 4.04 (SD = 0.92), with scores concentrated around 4, indicating consistent performance. Improved Retention demonstrates similarly high performance with a mean of 3.70 (SD = 0.95), though five participants reported significantly low scores (outliers at 1.0). Note that the data in Figure 2 represent aggregated responses from both courses. The boxplots reflect students' self-reported perceptions of learning efficiency, including time efficiency, retention, and interactivity. 'Improved retention' refers to how well students felt they remembered content, not an objective measure such as exam performance. Although a few outliers were observed, these responses may have come from students with different learning preferences or accommodations. Interactive Learning has a mean of 3.71 (SD = 0.97), with one notable outlier at 1.0. The small interquartile range (IQR) for all three metrics indicates tightly clustered results around the median (4.0), suggesting a generally positive and consistent outcome. These findings indicate that students generally found microlearning tools effective in terms of time efficiency, retention, and interactivity. Overall, the data highlights that microlearning materials are well-received, which supports their effectiveness in enhancing learning efficiency.

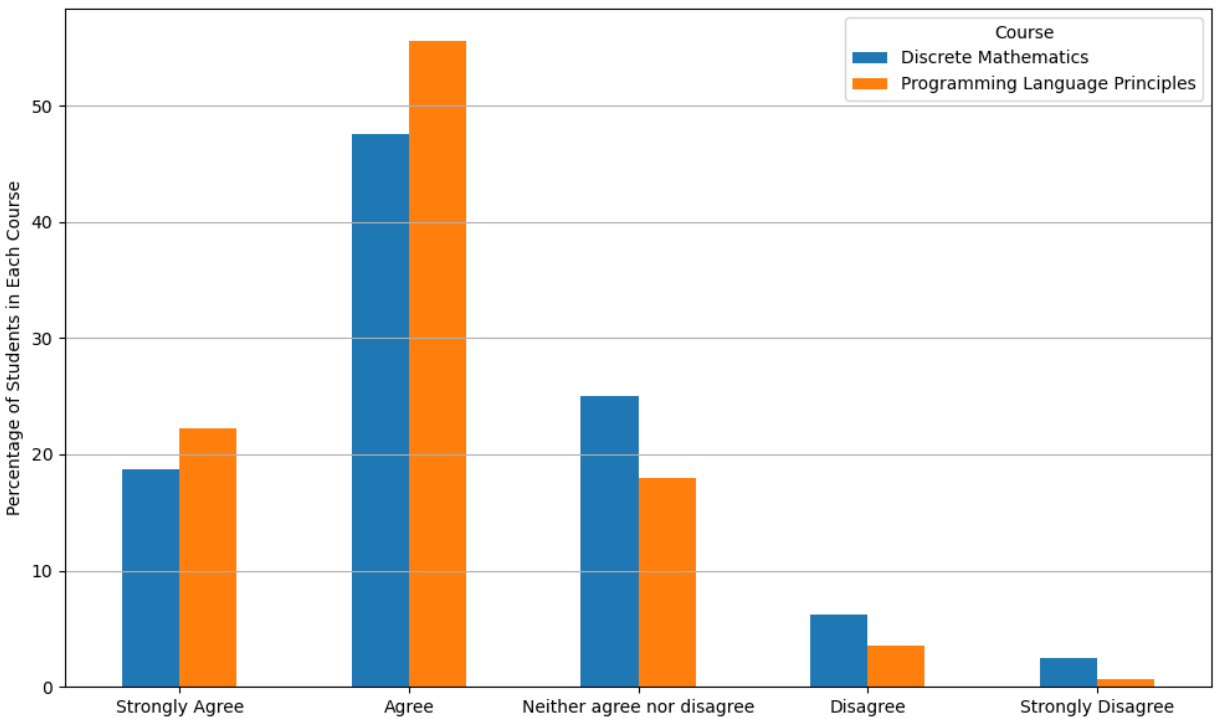

Figure 3: Students' feedback on the role of microlearning in improving engagement.

Figure 3 illustrates student perceptions of how well the course content fostered engagement in both courses. In Discrete Mathematics, approximately 67% of students either strongly agreed or agreed that the content effectively supported their engagement, compared to 80% in Programming Language Principles. Disagreement levels were low in both courses, with 8% (7 students) in Discrete Mathematics and 4% (12 students) in Programming Language Principles expressing dissatisfaction.

Figure 4 shows that over 65% of students in the Discrete Mathematics course either strongly agree or agree that microlearning serves as a good starting point for learning new topics. Similarly, more than 80% of students in the Programming Language Principles course share this view. Only a negligible number of students, 7% (6 students) in Discrete Mathematics and 5% (13 students) in Programming Language Principles, expressed disagreement or strong disagreement. These trends suggest that students in both courses perceive a more substantial alignment with the material provided as a foundational starting point.

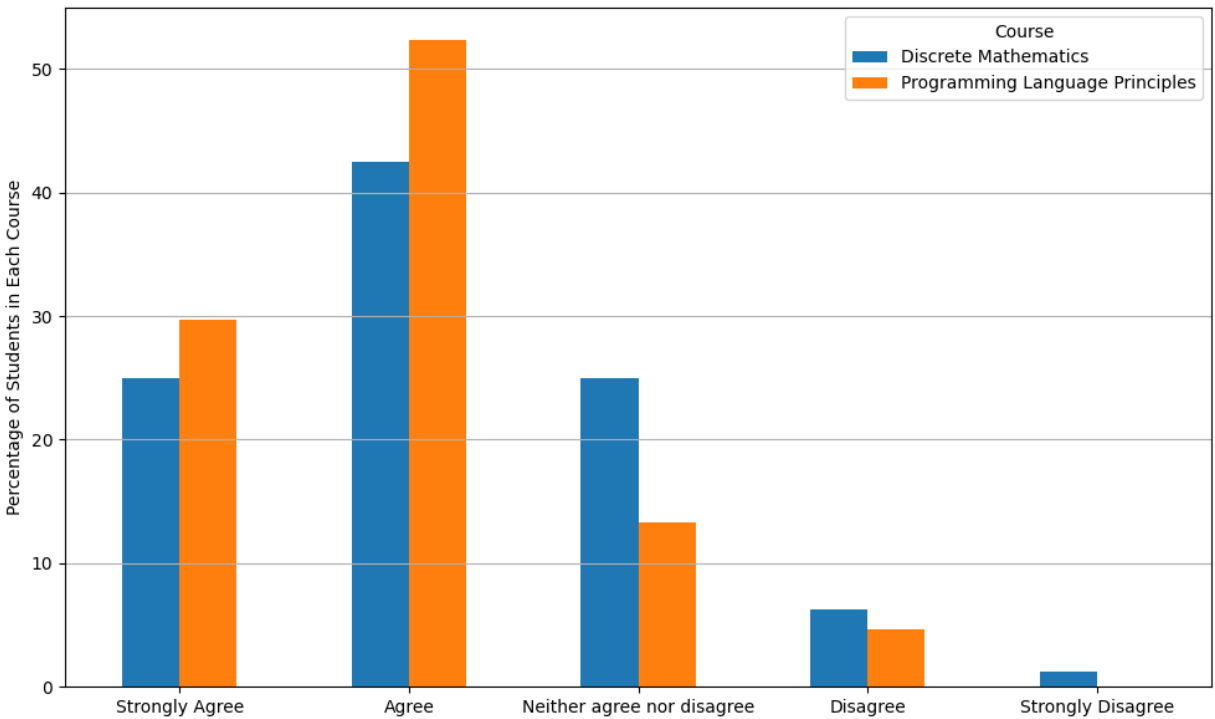

Figure 4: Students' feedback on the value of microlearning as a starting point for learning new topics.

*Addressing RQ#2*

Figure 5 indicates that over 35% of students in both courses reported never encountering incorrect information in the AI-generated content. However, a significant portion—more than 40% in Discrete Mathematics and over 50% in Programming Language Principles—indicated encountering incorrect information 1 to 3 times. Only a minimal number of students reported finding errors 8 to 10 times or more than 10 times, with responses ranging from 1 to 6 students, which are considered outliers, but for this study, it's important to consider each feedback. These findings underscore the need to address accuracy and consistency in AI-generated instructional materials, especially in programming courses, where even minor inaccuracies can significantly affect learning outcomes. The survey did not ask students to identify specific inaccuracies, so we could not determine whether these were unique issues or recurring errors found by multiple students. However, anecdotal review by instructors found fewer than 10 inaccuracies across all generated materials, suggesting that the 1–3 inaccuracies reported by students likely refer to the same errors. This highlights the importance of instructor-led instruction alongside AI-generated content, as solely relying on AI can pose risks to student learning. Although AI tools save time, it is essential for instructors to thoroughly review and revise AI-generated materials before incorporating them into course content. The results suggest that a hybrid approach, combining the efficiency of AI with the expertise of human instructors, is necessary to enhance the accuracy and effectiveness of AI-driven learning materials.

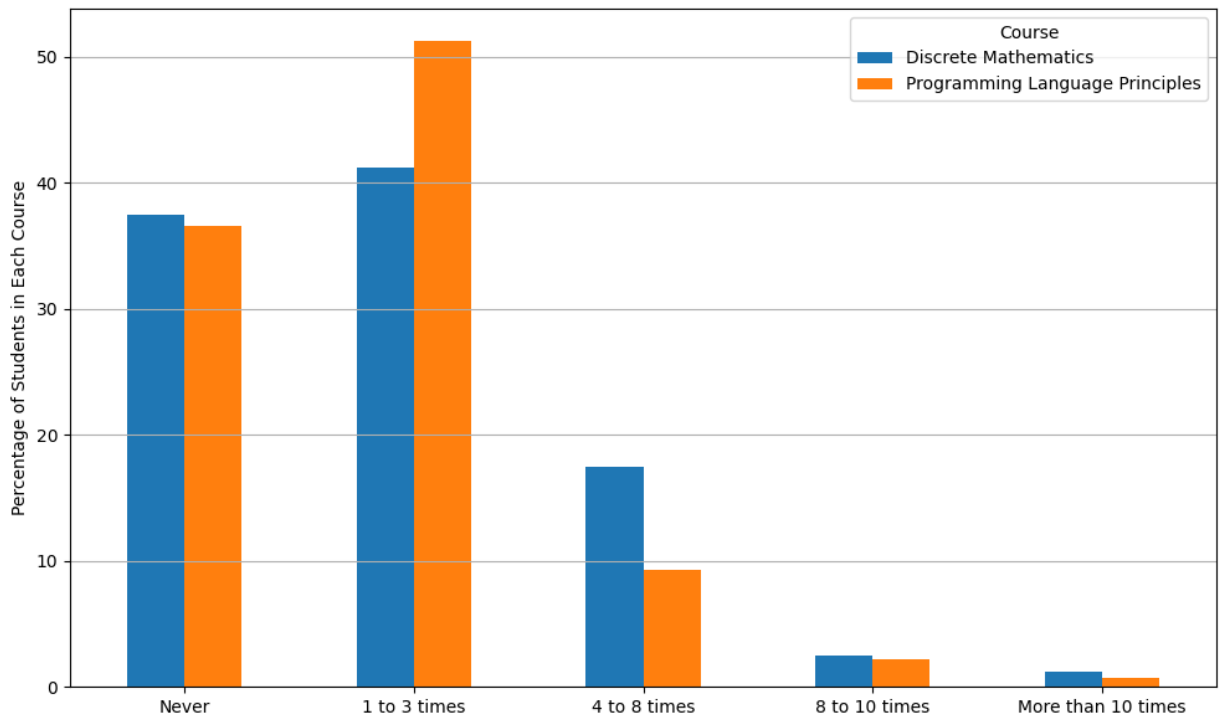

Figure 5: Students identified inaccuracies in AI-generated microlearning assessments.

Based on the analysis of the survey results depicted in Figure 6, there is a strong consensus regarding the relevance of the material taught in both courses. For Discrete Mathematics (Figure 6a), the relevance was with 76.5% responding "Yes," followed by 18.5% stating "Maybe," and a negligible 4.9% stating "No.".

Similarly, in Programming Language Principles (Figure 6b), 87.1% of respondents affirmed the relevance of the material ("Yes"), while 11.1% expressed uncertainty ("Maybe"), and only a small proportion (1.8%) indicated it was not relevant ("No"). These results suggest a positive perception of the course material's practical value among students.

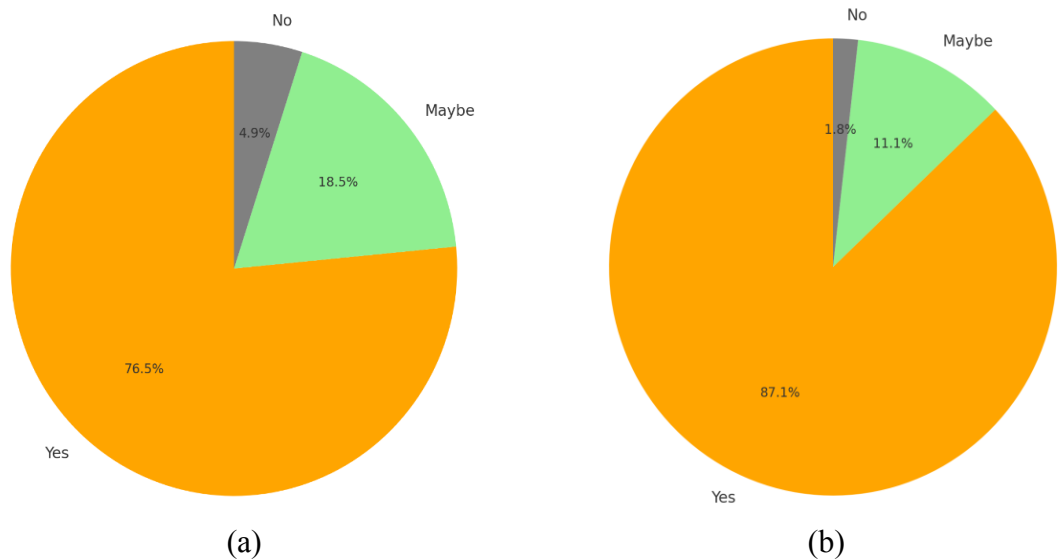

Figure 6: Students' feedback on the alignment of AI-generated microlearning content with actual course materials in (a) Discrete Mathematics and (b) Programming Language Principles.

**Discussion**

In 2005, Hug introduced a pivotal instructional framework for microlearning, which organizes it into seven dimensions: learning time, content, curriculum, form, process, medium, and type. These dimensions, which cover crucial aspects of microlearning, such as delivery formats, content structure, and instructional design, serve as a robust foundation for developing effective microlearning strategies [32]. While a variety of microlearning components have been explored and utilized in different contexts [22], [12], [28], [29], this study focuses on those commonly applied in educational settings.

Interactive quizzes and flashcards, two of the most versatile microlearning tools, are widely used in teaching for their ability to engage students and reinforce knowledge retention. Other strategies, such as mini summaries, which are prevalent in corporate training, seamlessly integrate microlearning with mobile learning (M-learning), allowing learners to access concise lessons conveniently through cell phone platforms [10]. Additionally, scenario-based activities, frequently employed in professional training and education programs, provide a platform for learners to develop decision-making and problem-solving skills in simulated environments. For instance, in nursing education, scenario-based training has proven effective in enhancing critical competencies by offering realistic clinical scenarios that foster critical thinking and improve clinical decision-making abilities [31].

While previous studies have highlighted the general benefits of microlearning and AI-driven tools, few have explicitly focused on evaluating key metrics—such as learning efficiency, engagement, and value as a starting point—across different academic contexts where AI-generated learning materials are employed. This research addresses that gap by analyzing the combined effects of content accuracy and relevance from the student's perspective, offering critical insights into optimizing AI-generated microlearning content to meet diverse student needs. By aligning educational technologies with pedagogical goals, this study contributes to the growing knowledge on enhancing learning outcomes through AI.

This study addressed Research Question 1 by capturing students' self-reported perceptions of learning efficiency, engagement, and the perceived value of microlearning as a starting point for learning. It is important to note that the study did not include direct assessments of learning outcomes—such as exam scores or performance metrics—and instead relied on subjective survey data to evaluate the effectiveness of the AI-generated materials. The box plot highlights positive student perceptions of microlearning, with a median score of 4 across Time Efficiency, Improved Retention, and Interactive Learning. Student engagement is a critical determinant of academic success, strongly influencing learning outcomes and retention rates. Research consistently shows active participation correlates with improved comprehension and performance [47]. Survey data comparing two courses—Programming Language Principles and Discrete Mathematics—revealed notable differences in how students engaged with AI-generated microlearning materials. These findings indicate a generally positive trend in engagement across both courses, with Programming Language Principles showing higher resonance, likely due to their interactive and practical elements. Exploring alternative content delivery methods may help enhance engagement in Discrete Mathematics.

In Programming Language Principles, 82.1% of students (52% agreeing and 30.1% strongly agreeing) affirmed that the microlearning materials effectively served as valuable starting points. The positive reception reflects the strong compatibility of microlearning with programming instruction. In contrast, discrete mathematics, rooted in abstract and theoretical concepts, showed a more mixed response. While 67.5% of students responded positively (42.5% agreeing and 25% strongly agreeing), 25% remained neutral, and 7.5% expressed disagreement. Higher neutrality and lower positive feedback may stem from the subject's complexity, which often requires deeper conceptual exploration and alternative delivery methods. While these findings are promising at the junior level, future research should examine whether similar benefits would be observed in freshman-level courses, where students may require more foundational support and different instructional scaffolding.

This disparity underscores the importance of tailoring microlearning strategies to meet the unique demands of each subject. For programming courses, structured and interactive tasks suit the material, while abstract subjects like Discrete Mathematics might benefit from additional context-driven explanations or real-world applications. These findings emphasize the need for instructional designers and educators to adapt microlearning content to align with course objectives while addressing students' cognitive preferences.

Adopting large language models (LLMs) in education has gained prominence due to their scalability and adaptability. However, their utility depends heavily on the accuracy and relevance of the content they produce which is related to our research question 2. This study found that students tolerate minor inaccuracies if the content is highly relevant. Survey results indicate that most students encountered inaccuracies only 1 to 3 times, and it is possible that these were the same inaccuracies for all students. This suggests that LLMs produce content with a generally acceptable level of accuracy for microlearning purposes. Furthermore, relevance ratings were particularly high for Programming Language Principles (87.1%), indicating that students value content utility despite minor errors.

Despite microlearning's recognized potential, studies reveal a gap in adoption: while 80.5% of teachers acknowledge its importance, only 18% frequently implement it [14]. This gap reflects the significant effort required to create and maintain microlearning materials. The proposed AI-driven method, by automating content creation and updating materials to reflect contemporary developments, offers a promising solution to this issue. Educators can feel relieved by the reduced time burden, enabling them to provide contextual insights and ensuring the material remains both accurate and relevant.

Integrating AI into education offers numerous benefits, including personalized learning experiences, improved learning outcomes, reduced educator workload, and innovative teaching methods. However, challenges persist, such as reduced human interaction, potential biases in AI systems, and the diminished role of educators. Generative AI tools like ChatGPT may also produce inaccurate content, necessitating thorough review by educators before dissemination to students. Balancing the strengths of AI with careful oversight will be critical to ensuring effective and ethical use in educational settings, making educators feel responsible and in control.

**Threat to Validity**

This study, while demonstrating the potential of AI-generated microlearning materials, is subject to certain threat to validity. Variations in grading policies across courses may have influenced student engagement levels and responses to the survey. For instance, graded microlearning activities in the Programming Language Principles course motivated higher participation than the ungraded Discrete Mathematics activities. Additionally, the self-reported nature of survey responses introduces a risk of bias, as students may have over- or under-reported their perceptions.

**Conclusion**

Core computer science courses are essential for obtaining a Computer Science and Engineering (CSE) degree. Our observations indicate that many students retake these courses to meet this criterion, indicating a need for supplementary assistance. High enrollment numbers pose challenges in managing these classes, and increased demand for help during office hours strains resources. Incorporating microlearning, such as quizzes and scenario-based learning, can enhance engagement and provide hands-on. Leveraging AI tools to create supplementary materials and innovative methods while reducing educator workload.

Our findings indicate that AI-generated microlearning materials significantly enhance student engagement and learning efficiency, particularly in application-driven courses like Programming Language Principles. However, the effectiveness varied in more abstract courses such as Discrete

Mathematics, suggesting the need for tailored approaches to different subject areas. Additionally, the accuracy of AI-generated content was generally high, with minimal inaccuracies reported, highlighting the reliability of large language models (LLMs) for educational purposes.

Overall, this research demonstrates the potential of AI-enhanced microlearning to transform educational practices, providing an adaptable and scalable framework for improving student outcomes and reducing educators' workload. These findings contribute valuable insights to the growing intersection of AI and education, paving the way for more innovative, student-centered learning solutions.

**Acknowledgments**

We thank the Leonhard Center in the College of Engineering at Penn State University for supporting this project. Their guidance and resources were crucial in completing our work successfully. We used ChatGPT [25] and Grammarly [15] to improve the sentence structure throughout the paper.

**References**

[1] T. I. Aldosemani, "Microlearning for Macro-outcomes: Students' Perceptions of Telegram as a Microlearning Tool," in *Digital Turn in Schools—Research, Policy, Practice*, Springer, Singapore, pp. 189–201, 2019.

[2] S. Ariyaratne, K. P. Iyengar, N. Nischal, N. C. Babu, and R. Botchu, "A comparison of ChatGPT-generated articles with human-written articles," *Skeletal Radiology*, vol. 52, no. 9, pp. 1755–1758, 2023.

[3] S. Bansal, "Textstat: Calculate statistical features from text," [Online]. Available: https://pypi.org/project/textstat/. [Accessed: Jul. 19, 2024].

[4] P. Basken, "Class attendance in US universities 'at record low,'" *Times Higher Education*, Dec. 6, 2023. [Online]. Available: https://www.timeshighereducation.com/news/class-attendance-us-universities-record-low. [Accessed: Jul. 19, 2024].

[5] I. Blagoev, G. Vassileva, and V. Monov, "From Data to Learning: The Scientific Approach to AI-Enhanced Online Course Design," in *Proc. 2023 Int. Conf. Big Data, Knowledge and Control Systems Engineering (BdKCSE)*, IEEE, pp. 1–5, 2023.

[6] L. Chen, P. Chen, and Z. Lin, "Artificial intelligence in education: A review," *IEEE Access*, vol. 8, pp. 75264–75278, 2020.

[7] M. C. Cortes-Albornoz, S. Ramírez-Guerrero, D. P. García-Guaqueta, A. VelezVan-Meerbeke, and C. Talero-Gutierrez, "Effects of remote learning during COVID-19 lockdown on children's learning abilities and school performance," 2023.

[8] N. Donthu, S. Kumar, D. Mukherjee, N. Pandey, and W. M. Lim, "How to conduct a bibliometric analysis: An overview and guidelines," *Journal of Business Research*, vol. 133, pp. 285–296, 2021.

[9] R. Flesch, "A new readability yardstick," *Journal of Applied Psychology*, vol. 32, no. 3, pp. 221–225, 1948.

[10] M. N. Giannakos, P. Mikalef, and I. O. Pappas, "Systematic literature review of e-learning capabilities to enhance organizational learning," *Information Systems Frontiers*, pp. 1–17, 2022.

[11] Y. Guo and H. Yu, "Exploration of education transformation and teacher literacy in the age of artificial intelligence," in *Proc. 5th Int. Workshop Artificial Intelligence and Education (WAIE)*, IEEE, pp. 38–42, 2023.

[12] C. Harrington, *Keeping Us Engaged: Student Perspectives (and Research-Based Strategies) on What Works and Why*. Taylor & Francis, 2023.

[13] The White House, "FACT SHEET: Biden-Harris Administration Announces Improving Student Achievement Agenda in 2024," Jan. 17, 2024. [Online]. Available: https://www.whitehouse.gov/briefing-room/statements-releases/2024/01/17/fact-sheet-biden-harris-administration-announces-improving-student-achievement-agenda-in-2024/. [Accessed: Jul. 19, 2024].

[14] C. Hu and Y. Liu, "Construction and Application of Micro Technology Platform in Applied Technology Universities under the Background of Internet Plus," *Journal of Physics: Conference Series*, vol. 1533, no. 1, 2020.

[15] Grammarly Inc., "Grammarly: Writing Assistant," [Online]. Available: https://app.grammarly.com/. [Accessed: Jul. 19, 2024].

[16] Kaltura Inc., "Kaltura: Online Video Platform," [Online]. Available: https://corp.kaltura.com. [Accessed: Jul. 19, 2024].

[17] J. Kim, P. J. Guo, D. T. Seaton, P. Mitros, K. Z. Gajos, and R. C. Miller, "Understanding in-video dropouts and interaction peaks in online lecture videos," in *Proc. 1st ACM Conf. Learning@Scale Conf.*, pp. 31–40, 2014.

[18] L. Kohnke, "Microlearning as a Teaching and Learning Approach," Springer Nature Singapore, Singapore, pp. 1–6, 2023. [Online]. Available: https://doi.org/10.1007/978-981-99-2774-6_1.

[19] T. Krasnova, A. Kouznetsova, Ovsyannikova, and A. Loginova, "Microlearning for Generation Z in the Foreign Language Classroom," in *EDULEARN23 Proceedings*, IATED, pp. 987–996, 2023.

[20] Y.-M. Lee, I. Jahnke, and L. Austin, "Mobile microlearning design and effects on learning efficacy and learner experience," *Educational Technology Research and Development*, vol. 69, no. 2, pp. 885–915, 2021.

[21] D. Leiker, A. R. Gyllen, I. Eldesouky, and M. Cukurova, "Generative AI for Learning: Investigating the Potential of Learning Videos with Synthetic Virtual Instructors," in *Artificial Intelligence in Education: Posters and Late Breaking Results, Workshops and Tutorials, Industry*

*and Innovation Tracks, Practitioners, Doctoral Consortium and Blue Sky*, Springer, Cham, pp. 523–529, 2023.

[22] K. Leong, A. Sung, D. Au, and C. Blanchard, "A review of the trend of microlearning," *Journal of Work-Applied Management*, vol. 13, no. 1, pp. 88–102, 2020.

[23] G. S. Mohammed, K. Wakil, and S. S. Nawroly, "The effectiveness of microlearning to improve students' learning ability," *International Journal of Educational Research Review*, vol. 3, no. 3, pp. 32–38, 2018.

[24] OpenAI, "OpenAI: Research and Development in Artificial Intelligence," [Online]. Available: https://openai.com/. [Accessed: Jul. 19, 2024].

[25] OpenAI, "ChatGPT: Language Model," [Online]. Available: https://openai.com/index/chatgpt/. [Accessed: Jul. 19, 2024].

[26] OpenAI, "Whisper: Automatic Speech Recognition System," [Online]. Available: https://openai.com/index/whisper/. [Accessed: Jul. 19, 2024].

[27] H. Ratnayake and C. Wang, "A Prompting Framework to Enhance Language Model Output," in *Proc. Australasian Joint Conf. Artificial Intelligence*, Springer, pp. 66–81, 2023.

[28] W. J. Rothwell, S. Singh, and J. Lee, *Accelerated Action Learning: Using a Hands-on Talent Development Strategy to Solve Problems, Innovate Solutions, and Develop People*. CRC Press, 2024.

[29] W. J. Rothwell, A. Zaballero, and F. Sadique, "Measuring the Return on Investment (ROI) in Technology-Based Learning," in *Trends and Issues in Instructional Design and Technology*, Routledge, Oxfordshire, England, UK, 2024. [Online]. Available: https://doi.org/10.4324/9781003502302.

[30] W. J. Rothwell, A. Zaballero, F. Sadique, and B. Bakhshandeh, *Revolutionizing the Online Learning Journey: 1,500 Ways to Increase Engagement*. CRC Press, 2024.

[31] M. Sadeghi, M. Nematollahi, J. Farokhzadian, Z. Khoshnood, and M. Eghbalian, "The effect of scenario-based training on the core competencies of nursing students: A semi-experimental study," *BMC Nursing*, vol. 22, no. 1, pp. 475–485, 2023.

[32] T. Hug, "Microlearning and narration," in *Fourth Media in Transition Conf.*, Cambridge, MA, 2005.

[33] J. Yin, T.-T. Goh, B. Yang, and Y. Xiaobin, "Conversation technology with micro-learning: The impact of chatbot-based learning on students' learning motivation and performance," *Journal of Educational Computing Research*, vol. 59, no. 1, pp. 154–177, 2021.

[34] P. F. Larrondo, B. M. Frank, and J. Ortiz, "Work-in-progress: Fine-tuning large language models for automated feedback in complex engineering problem-solving," in *Proc. 2024 ASEE Annual Conf. & Expo.*, 2024.

[35] I. Ivanov and H. Soliman, "Game of Algorithms: ChatGPT Implications for the Future of

Tourism Education and Research," *Journal of Educational Technology*, vol. 29, no. 2, pp. 45–62, 2023.

[36] P. Sridhar, A. Doyle, A. Agarwal, C. Bogart, J. Savelka, and M. Sakr, "Harnessing LLMs in curricular design: Using GPT-4 to support authoring of learning objectives," *arXiv preprint*, arXiv:2306.17459, 2023.

[37] K. D. Manning, J. O. Spicer, L. Golub, M. Akbashev, and R. Klein, "The micro revolution: Effect of Bite-Sized Teaching (BST) on learner engagement and learning in postgraduate medical education," *BMC Medical Education*, vol. 21, pp. 1–11, 2021.

[38] A. Cheung et al., "ChatGPT versus Human in Generating Medical Graduate Exam Multiple Choice Questions," *MedEd Journal*, vol. 35, no. 4, pp. 223–240, 2023.

[39] T. Hug, *Didactics of Microlearning*, Waxmann Verlag, 2007.

[40] R. Mayer, "Nine Ways to Reduce Cognitive Load in Multimedia Learning," *Educational Psychologist*, vol. 38, no. 1, pp. 43–52, 2003.

[41] R. Bartram et al., "Bite-Sized Teaching: Delivering Knowledge of Physical Health Issues in Mental Health Settings," *British Journal of Mental Health Nursing*, vol. 6, no. 6, pp. 265–271, 2017.

[42] K. Boumalek, A. El Mezouary, B. Hmedna, and A. Bakki, "Transforming Microlearning with Generative AI: Current Advances and Future Challenges," in *General Aspects of Applying Generative AI in Higher Education*, Springer, 2024, pp. 241-262. doi: 10.1007/978-3-031-65691-0_13.

[43] W. J. Rothwell, A. Zaballero, and S. Williams, *Experiencing Organization Development Practice through Case Stories and Role-Plays*, Cognella, 2025.

[44] A. Rof, A. Bikfalvi, and P. Marques, "Exploring learner satisfaction and the effectiveness of microlearning in higher education," *The Internet and Higher Education*, vol. 100952, 2024. Available: https://doi.org/10.1016/j.iheduc.2024.100952.

[45] D. Edge, S. Fitchett, M. Whitney, and J. Landay, "MemReflex: Adaptive flashcards for mobile microlearning," Microsoft Research Asia, 2012. [Online]. Available: https://dl.acm.org/doi/pdf/10.1145/2371574.2371641.

[46] C. Torgerson and S. Iannone, *Designing Microlearning*, Association for Talent Development, 2019.

[47] J. A. Gray and M. DiLoreto, "The effects of student engagement, student satisfaction, and perceived learning in online learning environments," *International Journal of Educational Leadership Preparation*, vol. 11, no. 1, p. n1, 2016.